%Paper: hep-th/9312181
%From: YVONNE@URHEP.PAS.ROCHESTER.EDU
%Date: Wed, 22 Dec 1993 14:41:06 -0500 (EST)

\magnification=1200
\baselineskip=20pt
\overfullrule=0pt
\def\bull{\vrule height .9ex width .8ex depth}% -.lex}

\centerline{\bf JORDAN TRIPLE SYSTEMS AND YANG-BAXTER EQUATION}

\vskip .5in
\centerline{by}
\vskip .5in
\centerline{Susumu Okubo}
\bigskip
\centerline{Department of Physics and Astronomy}
\centerline{University of Rochester}
\centerline{Rochester, NY 14627}

\vskip 1in

\noindent {\bf \underbar{Abstract}}

\medskip

Jordan as well as related triple systems have been used to find
several solutions of the Yang-Baxter equation, which are of
rational as well as trigonometric type.

\vskip 2in


\vfill\eject

\noindent {\bf 1. \underbar{Introduction}:}

The Yang-Baxter equation (hereafter referred to as YBE)
appears$^{1),2),3),4),5)}$ in many subjects ranging from exactly solvable
two-dimensional quantum field theory and exactly solvable two-dimensional
statistical mechanics problems to Hopf algebra, quantum group, and knot
theory.  In a series of papers$^{6),7),8),9)}$, the present author has
reformulated first the YBE as a triple product relation and then solved it
for triple systems called orthogonal and symplectic triple systems$^{6),7)}
$ in the first two papers.  The work has been then extended for super-space
verison in references 8 and 9.

The purpose of the present work is to find some solutions of YBE in terms
of Jordan triple system$^{10)}$ which turned out to require a slightly
different treatment to be described in this note.  However, in order to
make this paper as self-contained as possible, we will here briefly sketch
the basic formulation of our method.  Let $V$ be a $N$-dimensional vector
space over a field $F$, and we write
$$N = \ {\rm Dim}\ V \quad . \eqno(1.1)$$
Let $\underline{R} (\theta)$ for a parameter $\theta$ be a linear mapping
in $V \otimes V$, i.e.
$$\underline{R}
 (\theta) \ :\ V \otimes V \rightarrow V \otimes V \quad . \eqno(1.2)$$
For given basis vectors $e_1,\ e_2,\ \dots,\ e_N$ of $V$, we can then
expand
$$\underline{R} (\theta) e_a \otimes e_b = \sum^N_{c,d = 1} R^{dc}_{ab}
(\theta ) e_c \otimes e_d \eqno(1.3)$$
for scattering matrix elements $R^{dc}_{ab} (\theta)$.  Note the order of
indices $c$ and $d$ above.

Linear mappings $\underline{R}_{ij} (\theta)\ (i, j = 1, 2, 3)$ in $V
\otimes V \otimes V$ are now defined as usual by
$$\eqalignno{\underline{R}_{12} (\theta) e_a \otimes e_b
\otimes e_c &= \sum^N_{a^\prime , b^\prime = 1}
R^{b^\prime a^\prime}_{ab} (\theta) e_{a^\prime} \otimes
 e_{b^\prime} \otimes e_c \quad , &(1.4a)\cr
\underline{R}_{13} (\theta) e_a \otimes e_b
\otimes e_c &= \sum^N_{a^\prime , c^\prime = 1}
R^{c^\prime a^\prime}_{ac} (\theta) e_{a^\prime} \otimes
 e_b \otimes e_{c^\prime} \quad , &(1.4b)\cr
\underline{R}_{23} (\theta) e_a \otimes e_b
\otimes e_c &= \sum^N_{b^\prime , c^\prime = 1}
R^{c^\prime b^\prime}_{bc} (\theta) e_a \otimes
 e_{b^\prime} \otimes e_{c^\prime} \quad . &(1.4c)\cr}$$
In other words, $\underline{R}_{ij} (\theta)$  operates in the
$i$th and $j$th vector spaces just like $\underline{R} (\theta)$.  Suppose
that they satisfy the relation
$$\underline{R}_{12} (\theta) \underline{R}_{13} (\theta^\prime)
\underline{R}_{23} (\theta^{\prime \prime}) =
\underline{R}_{23} (\theta^{\prime \prime}) \underline{R}_{13} (\theta^\prime)
\underline{R}_{12} (\theta)  \eqno(1.5a)$$
with condition
$$\theta^\prime = \theta + \theta^{\prime \prime} \eqno(1.5b)$$
for parameters $\theta,\ \theta^\prime,\  {\rm and}\
 \theta^{\prime \prime}$.  The
equation (1.5a) with (1.5b) is called the Yang-Baxter equation (YBE).

If $\underline{R}_{ij} (\theta)$ satisfies another relation
$$\left[ \underline{R}_{12} (\theta), \underline{R}_{13} (\theta^\prime)
\right] + \left[
\underline{R}_{12} (\theta),
\underline{R}_{23} (\theta^{\prime \prime})
\right] + \left[  \underline{R}_{13} (\theta^\prime)
, \underline{R}_{23} (\theta^{\prime \prime})
 \right] = 0  \eqno(1.6)$$
instead of Eq. (1.5a) but with the condition Eq. (1.5b), it is called the
classical Yang-Baxter equation which appears$^{11)}$ in discussions of Lie
bi-algebras.

We can rewrite YBE's in a triple product form as follows$^{9)}$.
  Introduce a
bilinear nondegenerate form $<x|y>$ satisfying
$$<y|x>\ = \epsilon <x|y> \eqno(1.7)$$
in $V$, where $\epsilon$ assumes value 1 or $-$1.  We now set
$$g_{jk} = \epsilon g_{kj} = \ <e_j |e_k>\qquad
(j,k = 1,2,\dots,N) \eqno(1.8)$$
and write its inverse as $g^{jk}$ which satisfies
$$\sum^N_{\ell =1} g^{j \ell} g_{\ell k} = \sum^N_{\ell =1} g_{k \ell}
g^{\ell j} = \delta^j_k \quad . \eqno(1.9)$$
Introducing $e^j$ by
$$e^j = \sum^N_{k=1} g^{jk} e_k \quad , \quad e_j = \sum^N_{k=1}
g_{jk} e^k \quad , \eqno(1.10)$$
then any $x \ \epsilon\ V$ can be expanded as
$$x = \sum^N_{j=1} e_j <e^j |x>\ = \sum^N_{j=1} \ <x|e_j>e^j \quad .
\eqno(1.11)$$

In terms of the scattering matrix
 elements $R^{dc}_{ab} (\theta)$, we define two
$\theta$-dependent triple linear products $[x,y,z]_\theta$ and
 $[x,y,z]^*_\theta$ respectively by
$$\eqalignno{\big[ e^c,e_a,e_b \big]_\theta &= \sum^N_{d=1} e_d R^{dc}_{ab}
(\theta)  &(1.12a)\cr
\big[ e^d,e_b,e_a \big]^*_\theta &= \sum^N_{c=1}  R^{dc}_{ab}
(\theta)e_c  &(1.12b)\cr}$$
or alternatively by
$$R^{dc}_{ab} (\theta) =\ <e^d | \big[ e^c,e_a,e_b\big]_\theta>\ =\
<e^c | \big[e^d,e_b,e_a \big]^*_\theta> \eqno(1.13)$$
when we note
$$<e^d |e_c>\ = \delta^d_c \quad . \eqno(1.14)$$
Especially, it is more convenient to rewrite the second relation of Eq.
(1.13) as
$$<u |[x,y,z]^*_\theta>\ = \ <x|[u,z,y]_\theta> \eqno(1.15)$$
in a basis-independent notation for any $u,\ x,\ y,\ z\ \epsilon\ V$.  We
can now rewrite Eq. (1.3) as
$$\eqalign{\underline{R} (\theta) x \otimes y &= \sum^N_{j=1} e_j
\otimes [e^j, x,y]_\theta \cr
&= \sum^N_{j=1} \big[ e^j, y, x]^*_\theta \otimes e_j \cr}\eqno(1.16)$$
in a triple product notation when we note Eqs. (1.11) and (1.13).  The
Yang-Baxter equation (1.5a) can be then rewritten as a triple product
 equation
$$\eqalign{\sum^N_{j=1} \big[ &v,[u,e_j,z]_{\theta^\prime} ,
[e^j,x,y]_\theta \big]^*_{\theta^{\prime \prime}} \cr
&= \sum^N_{j=1} \big[ u,[v,e_j,x]^*_{\theta^\prime} ,
[e^j,z,y]^*_{\theta^{\prime \prime}} \big]_\theta \quad . \cr}\eqno(1.17)$$
Note that this relation is invariant under interchanges of
$$\theta \leftrightarrow \theta^{\prime \prime} \quad , \quad u
\leftrightarrow v \quad , \quad x \leftrightarrow z \quad , \quad
[x,y,z]_\theta \leftrightarrow [x,y,z]^*_\theta \quad . \eqno(1.18)$$
For the classical Yang-Baxter equation (1.6) we assume further the
additional condition of
$$<u|[x,v,y]_\theta>\ =\ <v|[y,u,x]_\theta> \eqno(1.19)$$
so that we will also have
$$<u|[x,v,y]^*_\theta>\ =\ <v|[y,u,x]^*_\theta> \eqno(1.19^\prime)$$
in view of Eq. (1.15).  We can then rewrite$^{9)}$ Eq. (1.6) as
$$\eqalign{\big[ u,&[v,z,x]^*_{\theta^\prime} ,y\big]_\theta +
\big[ u,x,[v,z,y]^*_{\theta^{\prime \prime}} \big]_\theta
+ \epsilon \big[ [x,u,v]^*_{\theta^\prime},z ,y \big]^*_{\theta^{\prime
\prime}} \cr
&= \big[ v,[u,x,z]_{\theta^\prime} ,y\big]^*_{\theta^{\prime \prime}} +
\big[ v,z,[u,x,y]_\theta \big]^*_{\theta^{\prime \prime}} +
 \epsilon \big[ [z,v,u]_{\theta^\prime},x ,y \big]_\theta\cr}\eqno(1.20)$$
which is also invariant under the interchange Eq. (1.18).  We remark here
that the condition Eqs. (1.19) and (1.19$^\prime$) will be satisfied for all
examples given in this paper.

We see from Eq. (1.17) that $[x,y,z]^*_\theta$ is also a solution of the
YBE, provided that $[x,y,z]_\theta$ is so.  This suggests the following.
Interchanging $[x,y,z]_\theta$ and $[x,y,z]^*_\theta$ in Eq. (1.16), we may
define $\underline{R}^* (\theta)\ :\ V \otimes V \rightarrow V
\otimes V$ by
$$\eqalign{\underline{R}^* (\theta)\ x \otimes y &= \sum^N_{j=1} e_j \otimes
\big[ e^j , x, y\big]^*_\theta\cr
&= \sum^N_{j=1} \big[ e^j , y, x\big]_\theta \otimes e_j \quad ,\cr}
\eqno(1.21)$$
as well as $\underline{R}^*_{ij} (\theta)\ (i, j=1,2,3)\ :\ V \otimes
 V \otimes V \rightarrow V \otimes V \otimes V$, analogously.  Defining the
permutation operation $P_{12} \ :\ V \otimes V \rightarrow V \otimes V$ by
$$P_{12}\  x \otimes y = y \otimes x \eqno(1.22)$$
then Eqs. (1.16) and (1.21) imply
$$\underline{R}^*(\theta) = P_{12} \underline{R} (\theta) P_{12}
\quad . \eqno(1.23)$$
Moreover, $\underline{R}^* (\theta) = \underline{R} (\theta)$ is equivalent
to the validity of
$$[x,y,z]^*_\theta = [x,y,z]_\theta \quad . \eqno(1.24)$$
Further, we may similarly define $\underline{P}_{ij}\ (i,j=1,2,3)\
:\ V \otimes V \otimes V \rightarrow V \otimes V \otimes V$ which
interchanges the $i$th and $j$th vector spaces.  Now we set
$$\underline{R}^*_{ij} (\theta) = \underline{P}_{ij}
\underline{R}_{ij} (\theta) \underline{P}_{ij}
\qquad (i,j=1,2,3) \eqno(1.25)$$
which satisfies another set of the YBE:
$$\underline{R}^*_{12} (\theta) \underline{R}_{13}^* (\theta^\prime)
\underline{R}^*_{23} (\theta^{\prime \prime}) =  \underline{R}^*_{23}
(\theta^{\prime \prime}) \underline{R}^*_{13}
(\theta^\prime) \underline{R}^*_{12}
(\theta) \quad .  \eqno(1.26)$$
The same fact also applies to the case of the calssical YBE.

For most of our applications, the condition (1.24) and hence
$\underline{R}^*_{ij}(\theta) = \underline{R}_{ij}
(\theta)$ is satisifed except for cases stated in Proposition 5 of section
4.

Although we can generalize$^{8),9)}$ our results for the case of $V$ being
a super-space, we will not discuss it in this note.  We organize our paper
as follows. We will first give the definition and examples of Jordan and
anti-Jordan triple systems in section 2, and solve in section 3  Eq.
(1.17) for the YBE
 as well as the classical YBE (1.20) in terms of some Jordan (and
anti-Jordan) triple products.  In section 4, we also find
additional solutions of
the YBE for a related and simpler triple system which may be called
associative triple system.  Some solutions satisfy both unitarity and
crossing symmetries$^{12)}$ so that they may correspond to some exactly
solvable two-dimensional quantum field theory.  Our solutions are of
rational as well as trigonometric types.

\medskip

\noindent {\bf 2. \underbar{Jordan and Anti-Jordan Triple Systems}}

Let $V$ be the $N$-dimensional vector space as in the previous section.
Let $x\ y\ z\ :\ V \otimes V \otimes V \rightarrow V$ be a
$\theta$-independent triple product in $V$ satisfying conditions:

\item{(i)} $z\ y\ x = \delta \ x\ y\ z$ \hfill (2.1)

\item{(ii)} $uv (xyz) = (uvx)yz - \delta x(vuy)z + xy(uvz)$
\hfill (2.2)

\noindent for a constant $\delta$ which can assume only two value 1 or
 $-$1.  The case of $\delta =1$ defines the standard Jordan triple
system$^{10)}$.  The other case of $\delta = -1$ is a little unfamiliar and
we may call it anti-Jordan triple system for a lack of better terminology.
We note that although Eq. (2.2) is a part of conditions for general
Freudenthal-Kantor triple systems$^{13),14),15),16)}$, the condition Eq.
(2.1) for $\delta = -1$ is not incorporated in general.

It is convenient for some time to introduce the left multiplication
operation $L_{x,y}\ :\ V \rightarrow V$ by
$$L_{x,y} z = x\ y\ z \quad . \eqno(2.3)$$
Then, Eq. (2.2) is rewritten as a Lie equation
$$\eqalign{\big[ L_{u,v},L_{x,y}\big] &= L_{uvx,y} - \delta L_{x,vuy}\cr
&= -L_{xyu,v} + \delta L_{u,yxv}\quad , \cr}\eqno(2.4)$$
where the second relation in Eq. (2.4) is a simple consequence of
interchanging $x \leftrightarrow u$ and $y \leftrightarrow v$ in the first
relation.  For a later purpose, we will state the following lemma.

\smallskip

\noindent \underbar{Lemma 1}

Let $x\ y\ z$ be either Jordan or anti-Jordan triple product.  Then, we
have
$$\eqalign{z(uyv)x &= xv(yuz) + zv(xuy) - yu(xvz)\cr
&= xu(zvy) + zu(yvx) - yv(zux) \quad . \cr}\eqno(2.5)$$

\smallskip

\noindent \underbar{Proof}

The first relation is a simple consequence of Eq. (2.2), i.e.
$$yu(xvz) = (yux)vz - \delta x(uyv)z + xv(yuz)$$
together with Eq. (2.1).  Then, interchanging $u \leftrightarrow v$ and
$x \leftrightarrow z$, and noting Eq. (2.1), the second relation follows
immediately. $\bull$

For all of our applications given in this paper, the vector space $V$
possesses always the bilinear nondegenerate form $<x|y>$ satisfying
 Eq. (1.7) as well as additional relations
$$\eqalignno{&<u|xvy>\ =\ <v|yux> &(2.6a)\cr
&<u|xvy>\ =\ \epsilon \delta <x|uyv>\ =\ \epsilon \delta
<y|vxu> &(2.6b)\cr}$$
which we will assume hereafter in this paper, where
$\epsilon$ is the same constant with values $+1$ or $-1$ as in Eq. (1.7).

As is well-known, we can always construct$^{10)}$ a Jordan triple
system from any Jordan algebra with the Jordan product $xy = yx$ by
$$x\ y\ z = x(yz) + z(yx) - (xz) y \quad .$$
We will give two examples of Jordan and anti-Jordan algebras satisfying
Eqs. (2.6) below, which will be used to obtain solutions of YBE in the next
section.

\smallskip

\noindent \underbar{Example 1}

Let $x \cdot y$ be an associative product in $V$, so that
$$x \cdot y \cdot z\ :\ = (x \cdot y)\cdot z = x \cdot
(y \cdot z) \eqno(2.7)$$
defines an associative triple product.  We next introduce the triple
product $x\ y\ z$ in $V$ by
$$x\ y\ z = x \cdot y \cdot z + \delta \ z \cdot y \cdot x \quad .
\eqno(2.8)$$
We can easily verify that Eqs. (2.1) and (2.2) are automatically satisfied.
 For $\delta =1$, this is not surprising, since the present case can
 then be
recovered from the special Jordan algebra with the Jordan product
$xy = x \cdot y + y \cdot x$.

In order to satisfy Eqs. (2.6), let $M_n$ be the vector space consisting of
all $n \times n$ matrices, i.e.
$$M_n = \{ x | x = n \times n \ {\rm matrix}\} \quad , \eqno(2.9)$$
which possesses the symmetric bilinear nondegenerate form
$<x|y>$ with $\epsilon =1$ by
$$<x|y>\ =\ <y|x>\ \equiv \ {\rm Tr}\ (x\cdot y) \quad . \eqno(2.10)$$
Then, the condition Eqs. (2.6) for $\epsilon=1$ are always satisfied in
$M_n$ for the triple product given by Eq. (2.8).  We now identify $V$ to be
various sub-spaces of $M_n$ as follows:

\vfill\eject

\smallskip

\noindent \underbar{Case (1a): Unitary Model $u(n)$}

$$V = M_n \quad , \quad N = n^2 \eqno(2.11)$$

\smallskip

\noindent \underbar{Case (1b): so(n) and sp(n) Models}

These cases are possible only for $\delta = 1$, i.e. the case of the Jordan
system.  We choose $V$ to be either of
$$\eqalignno{V^{(+)} &= \big\{ x| x\ \epsilon\ M_n , x^T = x \big\}
\quad , \quad N = {1 \over 2}\ n(n+1) \quad , &(2.12a)\cr
V^{(-)} &= \big\{ x| x\ \epsilon\ M_n , x^T = - x \big\}
\quad , \quad N = {1 \over 2}\ n(n-1) \quad . &(2.12b)\cr}$$
Here $x^T$ stands for the transpose matrix of $x$.

\smallskip

\noindent \underbar{Case (1c): Abelian Model}

$$V = V_0 \equiv \{ x | x\ \epsilon\ M_n , x =\ {\rm diagonal\  matrix}\}
\quad , \quad N = n
\quad  . \eqno(2.13)$$
This last case is also possible only for $\delta = 1$.  Moreover, since we
have $x \cdot y = y \cdot x$, we find
$$x\ y\ z = y\ x\ z = x\ z\ y = 2\ x \cdot y \cdot z \quad . \eqno(2.14)$$

\smallskip

\noindent \underbar{Example 2}

Let $J_\mu \ :\  V \rightarrow V$ be linear mappings in $V$ for any integer
$\mu$.  We assume that $J_\mu$'s satisfy the cyclic properties of the
period $n$ with multiplication table of

\item{(i)} $J_\mu J_\nu = J_{\mu + \nu}$ \hfill (2.15a)

\item{(ii)} $J_{\mu + n} = J_\mu$ \hfill (2.15b)

\item{(iii)} $J_0 = J_n = Id (=\ {\rm identity})$ \hfill (2.15c)

\noindent for any integer $\mu$ and $\nu$.  We assume moreover the validity
of
$$<x|J_\mu y>\ =\ <J_\mu x|y> \quad . \eqno(2.16)$$
Then, the triple product defined by
$$x\ y\ z = \sum^n_{\mu =1} \left\{ J_{n-\mu} x<y|J_\mu z>\ -\ J_{n- \mu} y<z
|J_\mu x>\ +\ J_{n-\mu} z<x|J_\mu y> \right\} \eqno(2.17)$$
can be shown to satisfy Eqs. (2.1), (2.2), and (2.6) with
$$\delta = \epsilon \quad . \eqno(2.18)$$
Moreover, the product obeys an additional identity
$$\eqalign{(yux)vz &- (yvz)ux \cr
&= \epsilon (yvz)xu - \epsilon (yxu)vz = \epsilon (yzv)ux
- \epsilon (yux)zv \quad . \cr}\eqno(2.19)$$
Interchanging $y \rightarrow z \rightarrow u \rightarrow y$, this yields
$$\left[ L_{u,v}, L_{x,y} + \epsilon L_{y,x} \right] = 0\eqno(2.20)$$
in terms of the left-multiplication operation $L_{x,y}$.  In other words,
$L_{x,y} + \epsilon L_{y,x}$ for any $x,\ y\ \epsilon\ V$ is a center
element of the Lie algebra defined by Eqs. (2.4).

For the application to be given in section 3, we have to impose further
restrictions of

$${\rm (i)}\ \  {\rm Tr}\  J_\mu \equiv \sum^N_{j=1} <e^j |J_\mu e_j>\ = N\
\delta_{\mu, 0} \hskip 2.5in \eqno (2.21a)$$

\item{(ii)} $N = 2n$ \hfill (2.21b)

\noindent by the reason to be given there.
  We can construct $J_\mu$'s satisfying all
these conditions as follows.  First let $J\ :\ V \rightarrow V$ be a linear
mapping in $V$ satisfying
$$<x|Jy>\ =\ <Jx|y> \quad , \quad J^n = Id \quad . \eqno(2.22)$$
Then, identifying
$$J_\mu = J^\mu \qquad (\mu{\rm -th\ power}) \quad , \eqno(2.23)$$
we see that all conditions of Eqs. (2.15) and (2.16) are automatically
obeyed.  In order to satisfy Eqs. (2.21), let
$$\omega = \exp \left( {2 \pi i \over n} \right) \eqno(2.24)$$
and choose $J$ to be a $N \times N$ diagonal matrix with diagonal entries
$(1,1,\omega, \omega, \omega^2, \omega^2, \dots, \omega^{n-1},$
 $\omega^{n-1}$) with $N= 2n$ for the case of $\epsilon = 1$.

\vfill\eject

\medskip

\noindent {\bf 3. \underbar{Solutions of YBE}}

We first start with the case of the classical YBE by showing the following
Proposition.

\smallskip

\noindent \underbar{Proposition 1}

Let $V$ be a triple system with triple product $x\ y\ z$ satisfying Eqs.
(2.4) and (2.6) with $\epsilon = \delta$.  Then,

\item{(i)} $[x,y,z]_\theta = {1 \over k \theta}\ y\ x\ z + B(\theta)
<x|y>z$\hfill (3.1a)

\noindent as well as

\item{(ii)} $[x,y,z]_\theta = {1 \over k \theta} \ (xyz-\epsilon yxz) +
 B (\theta) <x|y>z$ \hfill (3.1b)

\noindent are solutions of the classical YBE where $B(\theta)$ is an
arbitrary function of $\theta$ and $k$ is an arbitrary constant.  Moreover,
if the triple product obeys the additional condition of either

\item{(i)} $[L_{u,v} , L_{x,y} ]=0$ \hfill (3.2a)

\noindent or more weakly

\item{(ii)} $[L_{u,v} - \epsilon L_{v,u} , L_{x,y} - \epsilon
L_{y,x} ] =0$ \quad , \hfill (3.2b)

\noindent then

\item{(i)} $[x,y,z]_\theta = P(\theta) y\ x\ z +
 B(\theta) <x|y>z$ \hfill (3.3a)

\noindent and

\item{(ii)} $[x,y,z]_\theta = P(\theta) (xyz - \epsilon yxz) +
 B(\theta)<x|y>z$ \hfill (3.3b)

\noindent  for arbitrary functions $P(\theta)$ and $B(\theta)$ of
$\theta$ are the solutions of the classical YBE, respectively for
Eqs. (3.2a) or (3.2b).

\smallskip

\noindent \underbar{Remark 1}

Here, we need not assume the validity of Eq. (2.1) so that the conditions
(2.4) and (2.6)  will be satisfied$^{9)}$ not only by Jordan and
anti-Jordan triple systems but also by any orthogonal, symplectic, and some
Lie triple systems.  Further, the condition Eq. (3.2a) is satisfied by the
Abelian Jordan triple system of the example (1c) given by Eq. (2.13), since
then we have
$$\left[ L_{u,v}, L_{x,y} \right] z =
uv(xyz) - xy(uvz) = 0$$
identically when we note that $x\ y\ z = 2\  x \cdot y \cdot z$
 is a commutative
associative triple product by Eq. (2.14).  Moreover, if Eq. (3.2a)
holds, then $[x,y,z]_\theta = P(\theta) \ x\ y\ z + B(\theta) <x|y>z$ is
also a solution as we will see shortly.

\smallskip

\noindent \underbar{Proof}

We seek solutions with the ans\"atz of either Eq. (3.3a) or (3.3b) for some
functions $P(\theta)$ and $B(\theta)$ of $\theta$ to be determined.  We
first note that Eqs. (1.19) and (1.24) hold valid for both cases in view of
Eqs. (1.7) and (2.6) with $\epsilon \delta = 1$.  Inserting Eqs. (3.3) into
the classical YBE (1.20) and using Eq. (2.4), it is not hard to find that
they reduce respectively to

\item{(i)} $\{P(\theta) P(\theta^{\prime \prime}) - P(\theta)
 P(\theta^\prime) - P(\theta^{\prime \prime}) P(\theta^\prime)\}
[L_{x,u},L_{z,v}]y=0$ \hfill (3.4a)

\noindent or

\item{(ii)} $\{P(\theta) P(\theta^{\prime \prime}) - P(\theta)
 P(\theta^\prime) - P(\theta^{\prime \prime}) P(\theta^\prime)\}
[L_{x,u}- \epsilon L_{u,x} , L_{z,v} - \epsilon
 L_{v,z} ]y=0 \quad . $ \hfill (3.4b)

\noindent We note that any dependence upon $B(\theta)$ has disappeared from
Eqs. (3.4).  The solution of the function equation
$$P(\theta) P(\theta^{\prime \prime}) - P(\theta)
 P(\theta^\prime) - P(\theta^{\prime \prime}) P(\theta^\prime)
=0 \eqno(3.5)$$
with $\theta^\prime = \theta + \theta^{\prime \prime}$ is given by
 $P(\theta) = 1/k \theta$ for a constant $k$.  Also, Eqs. (3.4) are
identically satisfied under  the conditions of either Eq. (3.2a) or
(3.2b) for arbitrary $P(\theta)$.  These complete the proof of the
problem. $\bull$

For some special cases, we can do better.  Suppose that $V$ is now the
Jordan or anti-Jnordan triple system of the example 2 given by
Eq. (2.17), which satisfies the special relation of Eq. (2.20).  Then,
setting

\item{(i)} $[x,y,z]_\theta = P(\theta)\ x\ y\ z + B(\theta) <x|y>z
\quad ,$ \hfill (3.6a)

\noindent or

\item{(ii)} $[x,y,z]_\theta = P(\theta)(xyz + \epsilon yxz) +
B (\theta) <x|y>z \quad ,$\hfill (3.6b)

\noindent the analogues of Eqs. (3.4) now become

\vfill\eject

\item{(i)} $P(\theta) P(\theta^{\prime \prime}) \left[ L_{u,x},
L_{v,z} \right] y + P(\theta) P(\theta^\prime)
 \left[ L_{u,x}, \epsilon L_{z,v} \right] y $

\medskip

\item{   } $\qquad \qquad + P(\theta^\prime )P(\theta^{\prime \prime})
\left[ \epsilon L_{x,u}, L_{v,z} \right] y = 0$ \hfill (3.7a)

\noindent or

\item{(ii)} $P(\theta) P(\theta^{\prime \prime})
[L_{u,x} + \epsilon L_{x,u} , L_{v,z} + \epsilon L_{z,v}] y
 + P(\theta) P(\theta^\prime)
[L_{u,x} - \epsilon L_{x,u} , L_{v,z} + \epsilon L_{z,v}] y$

\medskip

\item{     } $\qquad \qquad + P(\theta^{\prime \prime}) P(\theta^\prime)
[L_{u,x} + \epsilon L_{x,u} , L_{v,z} - \epsilon L_{z,v}] y
 = 0 \quad .$   \hfill (3.7b)

\noindent Therefore, if the special relation Eq. (2.20) holds, then Eq.
(3.7b) is identically satisifed, while Eq. (3.7a) reduces to Eq. (3.4a).
Hence, we conclude that for the example 2 of Eq. (2.17),

\item{(i)} $[x,y,z]_\theta = {1 \over k \theta} \ x\ y\ z +
B(\theta) <x|y>z $ \hfill (3.8a)

\noindent as well as

\item{(ii)} $[x,y,z]_\theta = P(\theta) (xyz + \epsilon yxz) + B(\theta)
<x|y>z$ \hfill (3.8b)

\noindent are solutions of the classical YBE.

Similarly, from Eq. (3.7a) we see that Eq. (3.6a) is also a solution of the
classical YBE if we have $[L_{u,v},L_{x,y}]=0$.  Actually, Eq. (3.3a) is
also a solution of the standard YBE (1.17) under the conditions of Eqs.
(2.4), (2.6) and (3.2a) as we will see shortly in connection with the
Abelian model (1c) of Eq. (2.13) in Eq. (3.22a) below.  However, we will
not go into its detail in this paper.  Some examples of Lie triple systems
satisfying the condition Eq. (3.2a) will be given elsewhere.

Hereafter in what follows, we will consider solutions of the standard YBE
(1.17), assuming moreover the validities of Eqs. (2.1), (2.2), and
(2.6) with $\epsilon =1$.  We can find solutions for two cases of

\smallskip

\noindent \underbar{Case 1}

$$[x,y,z]_\theta = P(\theta) \ y\ x\ z + B(\theta) <x|y>z +
C(\theta) <z|x>y \quad , \eqno(3.9a)$$

\vfill\eject

\smallskip

\noindent \underbar{Case 2}

$$[x,y,z]_\theta = P(\theta) \ x\ z\ y + A(\theta) <y|z>x +
C(\theta) <z|x>y \quad . \eqno(3.9b)$$

\noindent The conditions Eqs. (1.19) and (1.24) are satisfied by Eq.
 (3.9a) while the same will also be satisfied by Eq. (3.9b) if
$\epsilon \delta = 1$.  Here, $P(\theta)$, $A(\theta)$, $B(\theta)$
and $C(\theta)$ are some functions of $\theta$ to be determined.

In contrast to the classical YBE, the solution for the standard YBE
requires some restrictions for the type of the Jordan or anti-Jordan triple
systems.  We will study the case 1 first, where we assume in addition the
following solvability ans\"atz:

$${\rm (i)}\ \  \sum^N_{j=1} (y e^j x)v e_j =
a\{ <x|v>y + \delta <y|v>x\}
+ b\ y\ v\ x \hskip 1.4in (3.10a)$$

\medskip

\item{(ii)} $\sum^N_{j=1} (ye^jx) v(zue_j) -
\sum^N_{j=1} (ye^jz)u(xve_j) = \alpha \{<v|x>zuy\  -
\ <u|z>xvy\}$

\item{   } $\qquad \qquad + \beta \{ <v|y>xuz \ -\ <u|y> zvx\  +\
<y|uzv>x$

\item{   } $\qquad \qquad - <y|vxu>z\} + \gamma \{
(yux)vz - (yvz) ux\}$ \hfill (3.10b)

\noindent for some constants $a,\ b,\ \alpha,\ \beta, {\rm and}\
\gamma$ to be specified shortly.

\smallskip

\noindent \underbar{Proposition 2}

Let $V$ be a Jordan or anti-Jordan triple system with
$\epsilon =1$ satisfying conditions Eqs. (2.1), (2.2) and (2.6) as well as
Eqs. (3.10).  Then,
$$[x,y,z]_\theta = P(\theta) \ y\ x\ z + B(\theta)
<x|y>z + C(\theta)<z|x>y \eqno(3.11)$$
for $P(\theta) \not= 0$ is a solution of the YBE (1.17) with
$${B (\theta) \over P (\theta)} = \delta \gamma + k \theta
\quad , \quad {C(\theta) \over P(\theta)} =
{\beta \delta \over k \theta} \eqno(3.12)$$
for an arbitrary constant $k$, provided that we have either

\item{(i)} $\alpha = \beta = 0 \quad ,$  \hfill (3.13a)

\noindent or

\item{(ii)} $\alpha = \beta \not= 0 \quad , \quad b = -2 \gamma \quad
, \quad a = 2 \beta$ \quad . \hfill (3.13b)

\noindent The solution moreover satisfies the unitarity condition
$$\eqalignno{&\underline{R}(\theta)\ \underline{R}(-\theta)
= f(\theta) \ Id &(3.14a)\cr
&f(\theta) = P(\theta) P(-\theta) \bigg\{ (a+\gamma^2) - (k \theta)^2
- {\beta^2 \over (k \theta)^2} \bigg\} &(3.14b)\cr}$$
where $Id$ in Eq. (3.14a) stands for the identity mapping in
 $V \otimes V$ as before.

\smallskip

\noindent \underbar{Remark 2}

All examples 1 and 2 of Jordan and anti-Jordan triple systems given in
 section 2 satisfy condition
 Eqs. (3.13).  As a matter of fact, values of $\alpha,$ $\beta$,
 $\gamma,$ $a,$ and $b$ for these examples are given by

\smallskip

\noindent \underbar{(1a) u(n) Case}
$$N = n^2 \quad , \quad \alpha = \beta = 1 \quad , \quad \gamma = 0
\quad , \quad a = 2 \quad , \quad b=0 \eqno(3.15a)$$

\smallskip

\noindent \underbar{(1b) so(n) and sp(n) Case}

$$N={1 \over 2} \ n(n \pm 1) \quad ,\quad \alpha = \beta = {1 \over 2}
 \quad ,
\quad \gamma = \mp {1 \over 2} \quad , \quad
a=1 \quad , \quad b =\pm 1 \eqno(3.15b)$$

\smallskip

\noindent \underbar{(1c) Abelian Case}

$$N=n \quad , \quad \alpha = \beta = \gamma = a = 0 \quad ,
\quad b=2 \eqno(3.15c)$$

\smallskip

\noindent \underbar{(2) Example 2}

$$N=2n \quad , \quad \alpha = \beta = 0 \quad , \quad \gamma = -n \quad ,
\quad a = 0 \quad , \quad b = 2n \eqno(3.15d)$$

\noindent where we assumed in addition the conditions Eqs.
(2.21).  $\bull$

\vfill\eject

\smallskip

\noindent \underbar{Proof of the Proposition 2}

We first note $[x,y,z]^*_\theta = [x,y,z]_\theta$ and insert Eq.
(3.11) into the YBE (1.17).  Using Eq. (2.5) of the lemma 1 as well as
other identities, we calculate for $\epsilon = 1$
$$\eqalign{\big[ v,&[u,e_j,z\big]_{\theta^\prime},
\big[ e^j,x,y\big] \big]_\theta \big]_{\theta^{\prime \prime}}
 - (u \leftrightarrow v, x
 \leftrightarrow z,
\theta \leftrightarrow \theta^{\prime \prime})\cr
&= K_1 \{(yux)vz - (yvz)ux\} + K_2 \{ <v|x>zuy \ -\ <u|z>xvy\}\cr
&\qquad + K_3 \{<v|y>xuz\   - <y|vxu>z \}
 - \hat K_3 \{<u|y>zvx\  - \ <y|uzv>x\}\cr
&\qquad + K_4 \{<z|u><y|v>x\  - \ <x|v><y|u>z\}\cr}\eqno(3.16)$$
where $K_j\ (j=1,2,3,4)$ are given by
$$\eqalignno{K_1 &= \delta \gamma P^{\prime \prime} P^\prime P +
 P^{\prime \prime} B^\prime P -  P^{\prime \prime} P^\prime B -
 B^{\prime \prime} P^\prime P &(3.17a)\cr
K_2 &= \delta \alpha P^{\prime \prime} P^\prime P -
 B^{\prime \prime} C^\prime P -  P^{\prime \prime} C^\prime B -
 \delta C^{\prime \prime} C^\prime  P \cr
&\qquad - \delta  P^{\prime \prime}
C^\prime C + \delta  C^{\prime \prime} P^\prime C - b  P^{\prime \prime}
C^\prime P &(3.17b)\cr
K_3 &= \delta \beta P^{\prime \prime} P^\prime P +
 C^{\prime \prime} P^\prime B -  C^{\prime \prime} B^\prime P &(3.17c)\cr
K_4 &= \delta a P^{\prime \prime} C^\prime P +
 B^{\prime \prime} C^\prime C +  C^{\prime \prime} C^\prime B -
 C^{\prime \prime} B^\prime C \quad . &(3.17d) \cr}$$
Here, we have set for simplicity
$$P = P(\theta) \quad , \quad P^\prime = P(\theta^\prime) \quad ,
\quad P^{\prime \prime} = P(\theta^{\prime \prime}) \eqno(3.18)$$
and similarly for $B(\theta)$ and $C(\theta)$.  Also, $\hat K_j \
(j=1,2,3,4)$ is the same function as $K_j$, except for the interchange of
$\theta \leftrightarrow \theta^{\prime \prime}$.  Note that we have $\hat
K_1 = K_1,$ $\hat K_2 = K_2$ and $\hat K_4 = K_4$ but $\hat K_3
\not= K_3$.  It is not hard to see that the conditions $K_1 = K_2 =
K_3 = K_4 = 0$ is satisfied by Eq. (3.12), provided that we have
$\alpha = \beta$, $\beta (2 \gamma + b) = 0$, and
$\beta (a -2\beta) =0$ or equivalently Eqs. (3.13).  Then, Eqs.
(3.14) can be readily verified.  This completes the proof of the
Proposition 2.  In order to show the validity of Eqs. (3.15) we note the
following.  Let us first consider the case of the $u(n)$ model
Eq. (2.11):

\vfill\eject

\smallskip

\noindent \underbar{Example (1a) u(n) Case}

Then, $e_j \ (1,2,\dots,N)$ are $n \times n$ matrices and we write their
matrix elements as $(e_j)_{\mu \nu}$ $(\mu , \nu =1,2,\dots,n)$.  The
completeness condition implies the validity of
$$\sum^N_{j=1} (e_j)_{\mu \nu} (e_j)_{\alpha \beta} =
\delta_{\mu \beta} \delta_{\nu \alpha} \quad . \eqno(3.19)$$
when we choose the basis to satisfy Tr$(e_j \cdot e_k) =
\delta_{jk}$ with $e^j = e_j$.
Let $X$ be an arbitrary $n \times n$ matrix, whose matrix elements are
$X_{\mu \nu}$.  Multiplying $X_{\nu \alpha}$ to both sides of Eq. (3.19),
this yields
$$\sum^N_{j=1} e_j \cdot X \cdot e_j = \ ({\rm Tr}\ X)
\  1 \eqno(3.19^\prime)
$$
in matrix notation, where 1 is the $n \times n$ unit matrix, and $A \cdot
B$ stands for the associative matrix product.

\smallskip

\noindent \underbar{Example (1b) so(n) and sp(n) Cases}

In this case, we must have $(e_j)^T = \pm e_j$
 in addition and accordingly the
relation Eq. (3.19$^\prime$) is now replaced by
$$\sum^N_{j=1} e_j \cdot X \cdot e_j = {1 \over 2}\ \left\{
({\rm Tr}\ X)\  1 \pm X^T \right\} \eqno(3.20)$$
for any arbitrary $n \times n$ matrix $X$, when we note
$$\sum^N_{j=1} (e_j)_{\mu \nu} (e_j)_{\alpha \beta} = {1 \over 2}\
\left\{ \delta_{\mu \beta} \delta_{\nu \alpha} \pm
\delta_{\mu \alpha} \delta_{\nu \beta}\right\}
\quad . \eqno(3.20^\prime)$$

\smallskip

\noindent \underbar{Example (1c) Abelian Case}

The corresponding relation is given by
$$\sum^N_{j=1} e_j \cdot X \cdot \cdot e_j = X \eqno(3.21)$$
for any diagonal matrix $X$.  Then, using these relations, we can find the
desired results Eqs. (3.15a), (3.15b) and (3.15c) after some calculations.

\vfill\eject

\smallskip

\noindent \underbar{Example 2}

Assuming the validity of Eq. (2.21a) but not yet Eq. (2.21b), a
straightforward calculation leads, for example, to
$$\eqalign{\sum^N_{j=1} (&y e^j x)v(zue_j) - (x \leftrightarrow z,u
\leftrightarrow v)\cr
&= - n \{(yux)vz - (yvz)ux\}\cr
&\qquad + (N-2n) \sum^n_{\alpha , \beta = 1} \{ <x|J_\alpha y><z|J_\beta
u>J_{-(\alpha + \beta)} v\cr
&\qquad - <z|J_\alpha y><x|J_\beta v>J_{-(\alpha + \beta)} u
\} \quad .\cr}$$
Therefore, only for the case of $N=2n$ as in Eq. (2.21b), the relation Eq.
(3.10b) can be satisfied with the result given in Eq. (3.15d). $\bull$

\smallskip

\noindent \underbar{Remark 3}

For some special cases, we can find more general solutions.  Consider the
Abelian case (1c) in Eq. (2.13).  In that case, we have identically
$$(yux)vz - (yvz)ux = 0$$
so that the condition $K_1=0$ is not needed to satisfy the YBE in
Eq. (3.16).  Solving $K_2 = K_3 = K_4 = 0$ with $\delta =1$, this yields
general solutions for three cases of
\item{(i)} $C(\theta) = 0,$ and both $P(\theta)$ and $B(\theta)$ being
arbitrary functions of $\theta$, \hfill (3.22a)

\item{(ii)} $B(\theta) = 0\ \ , \ \ {C(\theta) \over P(\theta)} = 2$
\hfill (3.22b)

\item{(iii)} $B (\theta) = 0\ \ ,\ \ {C(\theta) \over P(\theta)} =
{2 \over 1 - e^{k \theta}}$ \hfill (3.22c)

\noindent where $k$ is an arbitrary constant. $\bull$

Finally, we will consider the case 2 of Eq. (3.9b) with
 $\epsilon = 1$.  In this case, we impose the following analogue of Eq.
(3.10):

$${\rm (i)}\ \ \sum^N_{j=1} x e_j e^j = cx \quad ,
\hskip 3.5in  \eqno(3.23a)$$

$${\rm (ii)}\ \ \sum^N_{j=1} x(yze_j) e^j =
a <y|z>x\ +\ b\ xyz  \quad ,
\hskip 2.2in  \eqno(3.23b)$$

$${\rm (iii)}\ \ \sum^N_{j=1} x e^j (yze_j)  =
a <y|z>x\ +\ b\ xzy  \quad ,
\hskip 2.2in  \eqno(3.23c)$$

$$\eqalign{{\rm (iv)}\ \ \sum^N_{j=1} &v(e^j yx)(uze_j) - (u \leftrightarrow
v, x \leftrightarrow z) \hskip 2.7in \cr
&= \alpha \{ <y|zux>v\  -\  <y|xvz>u\  +\  <x|y>vzu\ \cr
&\qquad -\  <z|y>uxv\
+ <u|z>vxy \
 -\  <v|x>uzy\} + \gamma \{y(xuz)v\cr
&  \qquad - y(zvx)u + yx(uzv) -
yz(vxu)\} \cr}\eqno(3.23d)$$

\noindent where constants $a,\ b,\ c,\ \alpha,\ {\rm and}\
\gamma$ are given by

\smallskip

\noindent \underbar{Example (1a) u(n) Case}

$$\alpha =1 \quad , \quad \gamma =0 \quad , \quad a=2 \quad , \quad b=n
\quad , \quad
c=2n \eqno(3.24a)$$

\smallskip

\noindent \underbar{Example (1b) so(n) and sp(n) Case}

$$\alpha = {1 \over 2} \quad , \quad
\gamma  = \pm  {1 \over 4} \quad , \quad
a=1  \quad , \quad  b = {1 \over 2} \ (n \pm 1)
 \quad , \quad  c=n \pm 1 \eqno(3.24b)$$

\smallskip

\noindent \underbar{Example (1c) Abelian Case}

$$\alpha = \gamma = 0  \quad , \quad  a=0  \quad , \quad
b=1  \quad , \quad  c=1 \eqno(3.24c)$$

\vfill\eject

\smallskip

\noindent \underbar{Example (2)}

$$\alpha = 0  \quad , \quad  \gamma = {1 \over 2}\ n =
{1 \over 4} \ N  \quad , \quad  a=0  \quad , \quad
b = c = N \quad . \eqno(3.24d)$$
For Eq. (3.24d), we assumed the validities of Eqs. (2.21) again.
Unfortunately, it turns out now that the YBE (1.17) can be satisfied only
for $\gamma = 0$ with $\epsilon = \delta = 1$ i.e. only for either u(n)
case Eq. (3.24a) or the Abelian case Eq. (3.24c).  Here, we will discuss
only the u(n) case as in the following Proposition.

\smallskip

\noindent \underbar{Proposition 3}

Let $V$ be the Jordan triple system of the u(n) type as in Eq.
(2.11) with $\epsilon = \delta = 1$.  Then,
$$[x,y,z]_\theta = P(\theta) x\ z\ y + A(\theta) <y|z>x\ +
C(\theta) <z|x>y \eqno(3.25)$$
for $P(\theta) \not= 0$ offers solutions of the YBE (1.17) for
the following two cases:

\item{(i)} ${A(\theta) \over P(\theta)} = - {\lambda^2
e^{k \theta} - d \over \lambda (e^{k \theta}-d)}
\quad , \quad {C(\theta) \over P(\theta)} = -
{e^{k \theta} - \lambda^2 \over
\lambda (e^{k \theta}-1)}$ \hfill (3.26a)

\noindent where $d$ is either $\lambda^2$ or $-\lambda^4$ and $k$ is an
arbitrary constant, or

\item{(ii)} ${A(\theta) \over P(\theta)} = - \lambda \quad , \quad
{C(\theta) \over P(\theta)} = - {1 \over \lambda} \quad .$
\hfill (3.26b)

\noindent In both Eqs. (3.26a) and (3.26b), $\lambda$ is given by
$$\lambda = {1 \over 2}\ \left( n \pm \sqrt{n^2 -4}
\right) \quad . \quad \bull \eqno(3.27)$$

\smallskip

\noindent \underbar{Remark 4}

The first solution Eq. (3.26a) satisfies both unitarity and crossing
 symmetry relations:

$$\eqalignno{&\underline{R} (\theta) \underline{R} (-\theta) =
C(\theta) C(-\theta) Id &(3.28a)\cr
\noalign{\vskip 4pt}%
&{1 \over P (\overline \theta)} \ [y,x,z]_{\overline \theta}
= {1 \over P(\theta)} \ [x,y,z]_\theta &(3.28b)\cr}$$
where $\overline \theta$ in Eq. (3.28b) is related to $\theta$ by
$$\theta + \overline \theta = {1 \over k}\
\log d \quad . \eqno(3.29)$$
In view of these, the solution is likely related$^{12)}$ to some exactly
solvable two-dimensional quantum field theory.

\smallskip

\noindent \underbar{Proof of the Proposition 3}

We first calculate
$$\eqalign{[v,[&u,e_j,z]_{\theta^\prime} , [e^j,x,y]_\theta
]_{\theta^{\prime \prime}} - (u \leftrightarrow v,
x \leftrightarrow z, \theta \leftrightarrow \theta^{\prime \prime})\cr
&= K_1 \{<z|u>yxv\ -\ <x|v>yzu\} + K_2 \{ <x|y>uzv\cr
&\qquad -\ <y|xvz>u\}
 - \hat K_2 \{<z|y>vxu\ -\ <y|zux>v\}\cr
&\qquad + K_3 \{<x|y><z|u>v\ - \ <z|y><x|v>u\}\cr}\eqno(3.30)$$
where we have set
$$\eqalignno{K_1 &= P^{\prime \prime} P^\prime P +
n P^{\prime \prime} C^\prime P + P^{\prime \prime}
 C^\prime C + C^{\prime \prime} C^\prime P - C^{\prime \prime}
P^\prime C &(3.31a)\cr
K_2 &= P^{\prime \prime} P^\prime P + n P^{\prime \prime} P^\prime A +
P^{\prime \prime} A^\prime A + C^{\prime \prime} P^\prime A -
C^{\prime \prime} A^\prime P &(3.31b)\cr
K_3 &= 2\{ P^{\prime \prime} C^\prime P + P^{\prime \prime} P^\prime A +
A^{\prime \prime} P^\prime P\} +
2n \{ A^{\prime \prime} P^\prime A + A^{\prime \prime} C^\prime P +
P^{\prime \prime} C^\prime A\}\cr
&\qquad \qquad + A^{\prime \prime} A^\prime A + N A^{\prime \prime}
C^\prime A + A^{\prime \prime} C^\prime C +
C^{\prime \prime} C^\prime A - C^{\prime \prime} A^\prime
C \quad . &(3.31c)\cr}$$
Here, $\hat K_j \ (j=1,2,3)$ are again the same functions as $K_j$ except for
the interchange of $\theta \leftrightarrow \theta^{\prime \prime}$.  Note
that we have $\hat K_1 = K_1$ and $\hat K_3 = K_3$.  Solving equations
 $K_1 = K_2 = K_3 = 0$ with $N= n^2$, we find the result of Eqs. (3.26)
after some calculations.

We now see why $\gamma =0$ in Eq. (3.23d) is necessary for the solution,
since if $\gamma \not= 0$, then we must add a term
$$\gamma P^{\prime \prime} P^\prime P\{ y(xuz)v -y(zvx)u +
yx(uzv) - yz(vxu)\}$$
to the right side of Eq. (3.30).
We remark finally
 that the case of the 27-dimensional exceptional Jordan algebra
 is harder and will be examined in the future.

\smallskip

\noindent {\bf 4. \underbar{Solutions for Associative Triple Systems}}

We can generalize the results of section 3, if we directly deal with the
associative triple product instead of Jordan or anti-Jordan triple systems.
 Hereafter in this section, $x,\ y,\ z$ etc. stand for generic $n \times
 n$ matrices without any other constraint such as $x^T = \pm x$.
Especially, we have $N=n^2$ with the relation Eq. (3.19$^\prime$) and the
inner product given by Eq. (2.10), i.e.
$$<x|y>\ =\ <y|x>\ =\ {\rm Tr}\ (x \cdot y) \eqno(4.1)$$
as well as
$$\sum^N_{j=1} e^j \cdot X \cdot e_j = ({\rm Tr}\ X) \ 1 \eqno(4.2)$$
where $X$ stands for any $n \times n$ matrix, and $x \cdot y$ is the
associative matrix product.  We consider now a generalization of cases 1
and 2, respectively given by Eqs. (3.9a) and (3.9b) as follows:

\smallskip

\noindent \underbar{Case 1}

$$[x,y,z]_\theta = P_1 (\theta) z \cdot x \cdot y + P_2 (\theta)
y \cdot x\cdot z + B(\theta) <x|y>z\ + C(\theta) <z|x>y \eqno(4.3)$$

\smallskip

\noindent \underbar{Case 2}

$$[x,y,z]_\theta = P_1 (\theta) y \cdot z \cdot x + P_2 (\theta)
x \cdot z\cdot y + A(\theta) <y|z>x\ + C(\theta) <z|x>y
 \quad . \eqno(4.4)$$

\noindent Although we could have added $A(\theta)<y|z>x$ and $B(\theta)
<x|y>z$ to right sides of Eqs. (4.3) and (4.4) respectively, they do not in
general satisfy the YBE (1.17) so that we have simply dropped them.  For
the case 1, we have
$$[x,y,z]^*_\theta = [x,y,z]_\theta \quad . \eqno(4.5)$$
However, the same does not hold in general for the case 2, where we will
have
$$[x,y,z]_\theta^* = P_2 (\theta) y \cdot z \cdot x + P_1 (\theta)
x \cdot z\cdot y + A(\theta) <y|z>x\ + C(\theta) <z|x>y \eqno(4.6)$$
Inserting Eqs. (4.3)
 and (4.5) or (4.4) and (4.6) into the YBE (1.17) and using Eqs.
(4.1) and (4.2), we can find solutions of the Yang-Baxter equation.
However, since the calculation is somewhat lengthy and involved, we will
simply report here the final results without any detail of computations.

\vfill\eject

\smallskip

\noindent \underbar{Proposition 4}

Suppose that at least one of $P_1(\theta)$ and $P_2(\theta)$ is not
identically zero.  Then, the solutions of the YBE for the case 1, Eq.
(4.3), are given by

\item{(i)} $P_2 (\theta) = \lambda P_1 (\theta) \quad , \quad
{B(\theta) \over P_1 (\theta)} = k \theta \quad , \quad
 {C(\theta) \over P_1 (\theta)} = {\lambda \over k \theta}$
\hfill (4.7a)

\item{(ii)} $P_2 (\theta) = 0 \quad , \quad C(\theta) = 0 \quad , \quad
{B(\theta) \over P_1 (\theta)} = b + k \theta$ \hfill (4.7b)

\item{(iii)} $P_1 (\theta) = 0 \quad , \quad C(\theta) = 0 \quad , \quad
{B(\theta) \over P_2 (\theta)} = b + k \theta$ \hfill (4.7c)

\item{(iv)} $P_2 (\theta) = 0 \quad , \quad B(\theta) = 0 \quad , \quad
{C(\theta) \over P_1 (\theta)} = {1 \over  k \theta}
\quad {\rm or} \quad 0$   \hfill (4.7d)

\item{(v)} $P_1 (\theta) = 0 \quad , \quad B(\theta) = 0 \quad , \quad
{C(\theta) \over P_2 (\theta)} = {1 \over  k \theta}
\quad {\rm or} \quad 0 $ \hfill (4.7e)

\noindent where $\lambda,\ k,\ {\rm and}\ b$ are arbitrary constants.

\smallskip

\noindent \underbar{Remark 5}

If $\lambda = \delta = \pm 1$, then the solution Eq. (4.7a) reproduces the
u(n) type solution Eq. (3.12) in the Proposition 2 when we note
 $\gamma = 0$ and $\beta = 1$ by Eq. (3.15a).

Next, we will discuss the case 2 given by Eqs. (4.4) and (4.6).

\smallskip

\noindent \underbar{Proposition 5}

Suppose that at least one of $P_1 (\theta)$ and $P_2 (\theta)$ is not
identically zero in Eq. (4.4).  Then, the solutions of the YBE for the case
2 are given by

\item{(i)} $P_1 (\theta) = P_2 (\theta)$, and values of $A(\theta)$ and $C(
\theta)$
 being specified as in the Proposition 3 with $P(\theta) \equiv
 P_1 (\theta) = P_2 (\theta)$.

\item{(ii)} $P_2 (\theta) = 0 \quad , \quad {C(\theta)
\over P_1(\theta)}  = {n \over e^{k \theta} -1} \quad , \quad
{A(\theta) \over P_1 (\theta)} = {n e^{k \theta} \over
(n^2 -1)-e^{k \theta}}$   \hfill (4.8a)

\item{(iii)} $P_2 (\theta) = 0 \quad , \quad {C(\theta) \over
P_1 (\theta)} = -n  \quad , \quad
A(\theta) =0$   \hfill (4.7b)

\item{(iv)} $P_2 (\theta) = 0 \quad , \quad C(\theta) = 0 \quad , \quad
{A(\theta) \over P_1 (\theta)} = -n$   \hfill (4.7c)

\noindent as well as solutions obtained by interchanging $P_1(\theta)
\leftrightarrow P_2(\theta)$.  Here, $k$ is again an arbitrary constant.
 $\bull$

\smallskip

\noindent \underbar{Remark 6}

We note that the Yang-Baxter equation (1.17) is invariant under the
interchange of
$$[x,y,z]_\theta \leftrightarrow [x,y,z]^* \eqno(4.9)$$
as we noted already in section 1.  This is the reason why we can find also
solutions of the YBE in the Proposition 5 by interchanging $P_1 (\theta)
\leftrightarrow P_2 (\theta)$.  $\bull$

\smallskip

\noindent \underbar{Remark 7}

The solution Eq. (4.8a) also obeys both unitarity and crossing symmetric
relations
$$\eqalignno{&\underline{R} (\theta) \underline{R}^* (-\theta) =
\underline{R}^* (-\theta) \underline{R}(\theta) =
C(\theta) C(-\theta) \ Id \quad , &(4.10)\cr
\noalign{\vskip 4pt}%
&{1 \over P_1 (\overline \theta)} \ [y,x,z]_{\overline \theta} =
{1 \over P_1 (\theta)}\ [x,y,z]_\theta &(4.11)\cr}$$
where $\overline \theta$ is related to $\theta$ by
$$\theta + \overline \theta = {1 \over k}\ \log (n^2 -1)
\quad . \eqno(4.12)$$
Here, $\underline{R}^*(\theta)$ is defined by Eq. (1.23). $\bull$

In ending this paper, we simply remark that G\"unaydin and coworkers$^{17)}$
 as well as Truini and Biedenharn$^{18)}$
 have used Jordan-triple systems and Jordan-pair triple systems for
 some other physical problem.

\smallskip

\noindent {\bf \underbar{Acknowledgements}}

This paper is supported in part by the U.S. Department of Energy Grant

\noindent DE-FG-02-91ER40685.

\vfill\eject

\baselineskip=18pt

\noindent {\bf \underbar{References}}

\item{1.} \underbar{Yang-Baxter Equation in Integrable Systems}, ed. by
 M. Jimbo, World Scientific, Singapore (1989).

\item{2.} \underbar{Integrable Systems and Quantum Groups}, ed. by M.
Carfora, M. Martelli, and A. Marzuoli, World Scientific, Singapore
(1992).

\item{3.} \underbar{Braid Group, Knot Theory and Statistical Mechanics},
ed. by C. N. Yang and M. L. Ge, World Scientific, Singapore (1989).

\item{4.} L. H. Kauffman, \underbar{Knots and Physics}, World Scientific,
Singapore (1991).

\item{5.} Y. I. Manin, \underbar{Quantum Groups and Non-Commutative
Geometry}, University of Montr\'eal Press, Montr\'eal (1988).

\item{6.} S. Okubo, Jour. Math. Phys. {\bf 34}, 3273 (1993).

\item{7.} S. Okubo, Jour. Math. Phys. {\bf 34}, 3292 (1993).

\item{8.} S. Okubo, University of Rochester Report UR-1312 (1993), to
appear in the Proceedings of the 15th Montr\'eal-Rochester-Syracuse-Toronto
Meeting for high energy theories.

\item{9.} S. Okubo, University of Rochester Report UR-1319 (1993)
unpublished.

\item{10.} N. Jacobson, \underbar{Structure and Representations of Jordan
Algebras}, Amer. Math. Soc. (1971).

\item{11.} V. G. Drinfeld, Quantum Groups: in the Proceedings of the
International Congress of Mathematicians, Berkeley, 1986.

\item{12.} A. B. Zamolodchikov and A$\ell$. B. Zamolodchikov, Ann. Phys.
{\bf 120}, 253 (1979).

\item{13.} K. Yamaguchi, in S\=urikagaku K\=oky\=uroku {\bf 308},
University of Kyoto, Inst. Math. Analysis (1977) in Japanese.

\item{14.} I. Bars and M. G\"unaydin, Jour. Math. Phys. {\bf 20}, 1977
 (1979).

\item{15.} K. Yamaguchi, Bull. Fac. Sch. Educ. Hiroshima University
 {\bf 6} (2), 49 (1983).

\item{16.} N. Kamiya, Jour. Algebra {\bf 110}, 108 (1987).

\item{17.} M. G\"unaydin and S. Hyun, Nucl. Phys. {\bf B373}, 688
(1992) and references quoted therein.

\item{18.} P. Truini and L. C. Biedenharn, J. Math. Phys.
{\bf 23}, 1327 (1982).

\end